# Classifications of Innovations
# Survey and Future Directions

Mario Coccia

*(Ceris-Cnr, Italia & Max-Planck Institute of Economics, Germania)*

National Research Council of Italy, Ceris-Cnr
Via Real Collegio, 30
10024 Moncalieri (Torino) – Italy
Tel.: +39.011.6824.925; Fax: +39.011.6824.966; email: m.coccia@ceris.cnr.it

ABSTRACT. The purpose of this paper is to focus on similarity and/or heterogeneity of taxonomies of innovation present in the economic fields to show as the economic literature uses different names to indicate the same type of technical change and innovation, and the same name for different types of innovation. This ambiguity of classification makes it impossible to compare the various studies; moreover the numerous typologies existing in the economics of innovation, technometrics, economics of technical change, management of technology, etc., have hindered the development of knowledge in these fields. The research presents also new directions on the classification of innovation that try to overcome these problems.

KEYWORDS: Classifications, Taxonomy, Technical change, Product, Innovation Patterns, Management of Technology, Economics of innovation

JEL CODES: B11, B12, B20, B41, O30

ACKNOWLEDGEMENTS. I am grateful to Secondo Rolfo (*Ceris-Cnr, Italia*), Ian McCarthy (*SFU Business, Simon Fraser University, Vancouver; Canada)* for useful suggestions to a preliminary draft of this paper. Any errors are my sole responsibility.

CONTENTS

## INTRODUCTION

The capacity to innovate is an important strategic option for many firms and countries. It is also a central and enduring research theme for academics, which have spent the last 70 years defining, explaining and measuring innovation in its many forms. A popular and fundamental approach that has accompanied these studies is the classification of innovation, which is both a process (to classify) and an output of the process (a classification). The classification provides models for ordering, labeling, and articulating knowledge about the diversity of innovations. Classification helps us to arrange and structure our knowledge in a way that is more fruitful and transferable that a simple list of descriptions.

The classifications of technical change and innovation, and its interpretation, remains one of the most difficult problems for scholars to analyze, due to the several variables involved and because the innovation can have different causes of origin.

As classification is a common process in the physical, life and social sciences: the result is a diverse range of interpretations and frequent misuse of classification terms, theories and methods. Although the words "category" and "taxonomy" are almost synonyms, they are very different in age. As early as 2,300 years ago, the father of all taxonomies, Aristotle, often used, the word "Kathegoría". The word taxonomy is, on the contrary, a recent one, dating to the first half of the eighteenth century. Several scholars, including Linnaeus, to classify minerals and animal and botanical species used it. The scientists and philosophers of the Enlightenment introduced this neologism by recovering an ancient Greek word (táxon, arrangement, array) and associating it to nómos. Since then the word has been very successful and is still used in natural sciences to classify species, minerals and the phenomena. It should be noted that the term flourished in the natural sciences a century before Charles Darwin proposed his theory of evolution, though in more recent times taxonomies have tried to describe and explain the static characteristics of objects as well as their evolving patterns. Over the last decades, the word has also been imported in social sciences. Taxonomies are meant to classify phenomena with the aim of maximizing the differences among groups. The term taxonomy refers to a branch of systematics concerned with the theory and practice of producing classification schemes. Thus, constructing a classification is a taxonomic process with rules on how to form and represent groups (taxa), which are then named (nomy). The social sciences have two general approaches to classification: the empirical and theoretical. The principal difference between these two social science approaches is the stage at which a theory of differences is proposed and evidence then sought to validate the theory (Warriner, 1984; Rich 1992; Doty and Glick, 1994). Theoretical classifications in the social sciences begin by developing a theory of differences which then results in a classification of organizational types, known as a typology. Only when the classification has been proposed is a decision made as to where an entity belongs in the classification. With the empirical approach, social science classifications begin by gathering data about the entities under study. The data are then processed using statistical methods (numerical taxonomy) to produce groups according to the measures of similarity and statistical techniques used. Thus the overall aim is to use data to construct the classification, instead of supporting it, but in reality, data are seldom collected without an expectation about what they will reveal or validate. While, for example, a "taxonomy" is considered useful, if it is able to reduce the complexity of the population studied into easily recallable macro-classes, the “classifications” are often highly desegregated, both in natural and social sciences (Archibugi, 2001). Although the term classification has been used throughout this paper, there is no agreement about the general use of the term. Classification as an output (a product of the process of classifying) deals with how groups and classes of entities will be arranged, in accord with the taxonomic approach used (McKelvey, 1982). It is a framework (e.g. a matrix, a table, a dendrogram, etc.) for ordering and representing, regardless of whether a theoretical or empirical approach is used. The terms classification scheme or classification system are often used to distinguish and identify classification as an output. Examples of such schemes and systems include the Linnaean System of nomenclature, the Periodic Classification of chemical elements, the Mercalli scale, the Dewey Decimal Classification System for organizing books and other bibliographic items and the North American Industrial Classification (NAIC) and Standard Industrial Classification (SIC) systems for naming and organizing industry sectors.

The purpose of this paper is to focus on similarity and/or heterogenity of taxonomies of innovation present in the economic fields to show as the economic literature uses different names to indicate the same type of technical change and innovation, and the same name for different types of innovation. The taxonomies of innovation can be divided in two sets of analysis (section 1): economics of innovation and management of technology. Section two presents a discussion and the new directions in the classification of the technical change and innovation that try to overcome the pervious problems.

## 1. CLASSIFICATION OF TECHNICAL CHANGE AND INNOVATION INTENSITY: A LITERATURE REVIEW

Souder and Shrivastrava (1985) said “we can’t begin to make decisions about technology until we understand it, and we can’t begin to really understand it until we can measure it”.

Garcia and Calantone (2002) state that innovations are frequently classified in taxonomies in order to identify their innovation characteristics and the degree of innovativeness involved. According to Durand (1992) four different perspectives can be adopted in order to analyze the intensity and the significance of technical change: 1) Technological input: technical novelty or scientific merit; 2) Competence throughput: new requirements on the competencies (resources, skills and knowledge), transilience (Abernathy and Utterback, 1985); 3) Perception of the market: market novelty, new functions proposed to customers; 4) Strategic output: impact on the competitive position of the firms.

Empirical studies classify innovations in two fields: 1) at the macro level, the characteristics of innovation that are new to the world, market or sector are considered (Maidique and Zirger, 1984; Lee and Na, 1994). In this case, the innovativeness is based on factors that are exogenous to the firm, such as the familiarity of an innovation to the world and the industry or the creation of new competitors due to the introduction of new innovations; 2) at the micro level, innovation is new to firm or to the consumer (More, 1982). Some researchers use both yardsticks (Ali *et al*., 1995; Cooper, 1979; Cooper and de Brentani, 1991). The innovation classifications can be divided according to two fields of study. These two perspectives show different characteristics of innovation will be as described in the following sections.

### *1.1 Taxonomies of innovation in economics of innovation*

After Schumpeter (1939), according to whom technical knowledge is acquired both through invention and innovation, economists identified several kinds of innovations within technical change (Mensch, 1979; Priest and Hill, 1980; Archibugi and Santarelli, 1989; Durand, 1992; Dosi, 1988; Clark, 1985; Freeman, 1984).

Pavitt (1984) classified innovation according to the firms that generate it, identifying four sectorial taxonomies. Pavitt intended the taxonomy to describe the behavior of innovating firms, to predict their actions and to suggest a framework for policy analysis. Taxonomy was composed of four main categories:

- The first was *supplier dominated* firms active in traditional industries such as clothing and furniture (i.e. firms which innovate by acquiring machinery and equipment).
- The second was *specialized suppliers* of capital goods and equipment who live in symbiosis with their customers.
- The third was *science-based firms* born to exploit new scientific discoveries in fields such as electronics, chemicals, pharmaceuticals and aerospace, where the main source of knowledge is associated with in-house R&D laboratories.
- The fourth was *scale-intensive* firms active in mass production industries.

In subsequent versions Pavitt has added another category to classify the emerging information-intensive firms, which have their main source of technological accumulation in the advanced processing of data and are typical in sectors or industries such as banking, retailing, internet, software, and so on. According to Archibugi (2001) this has led to the disappearance of one of the former categories: namely specialized supplier firms. According to Pavitt's latest thoughts, these firms are somehow forced to become information-intensive or scale-intensive or to become non-innovative: "We have also excluded a 'supplier dominated' trajectory since ... it leaves accumulated technological skills and strategic initiative with suppliers. Firms intending to move from this position try to adopt either scale-intensive strategies (e.g. certain textile firms), or information-intensive strategies (e.g. certain retailing firms)" (Pavitt *et al*., 1989, p. 96-97). Archibugi *et al*. (1991), state that supplier-dominated firms have a distinctive and significant technological trajectory and can be equally innovative by acquiring machinery and capital equipment.

Freeman *et al*. (1982), Freeman and Soete (1987) categorize various types of technical change and distinguish among:

- INCREMENTAL INNOVATIONS. *These occur more or less continuously in any industry or service activity, although at a varying rate in different industries and over different time periods. They may often occur, not so much as the result of formal research and development activity, but as the outcome of inventions and improvements suggested by engineers and others directly engaged in the production process, or as a result of initiatives and proposals by users. Many empirical studies have confirmed their great importance in improving the efficiency in use of all factors of production, for example Townsend's (1981) study of the Anderton shearer loader in the British coalmining industry. They are particularly important in the follow-through period after a radical breakthrough innovation and frequently associated with the scaling up of plant and equipment and quality improvements to products and services for a variety of specific applications. Although their combined effect is extremely important in the growth of productivity, no single incremental innovation has dramatic effects, and they may sometimes pass unnoticed and unrecorded. However, their effects are apparent in the steady growth of productivity, which is reflected in input-output tables over time by major changes in the coefficients for the existing array of products and services* (Freeman *et al.*, 1982).

- RADICAL INNOVATIONS. *These are discontinuous events and in recent times is usually the result of a deliberate research and development activity in enterprises and/or in university and government laboratories. They are unevenly distributed over sectors and over time. Freeman and Soete's research did not support the view of Mensch (1979) that their appearance is concentrated particularly in periods of deep recessions. They would agree with Mensch that, whenever they may occur, they are important as the potential springboard for the growth of new markets, or in the case of radical process innovations, such as the oxygen steelmaking process, of big improvements in the cost and quality of existing products. Over a period of decades a radical innovation, such as nylon or the contraceptive pill, may have fairly dramatic effects, but in terms of their economic impact they are relatively small and localized, unless a whole cluster of radical innovations are linked together in the rise of entire new industries and services, such as the synthetic materials industry or the semiconductor industry. Strictly speaking, at a sufficiently disaggregative level, radical innovations would constantly require the addition of new rows and columns in an input-output table. But in practical terms, such changes are introduced only in the case of the most important innovations and with long time-lags, when their economic impact is already substantial* (Freeman *et al.*, 1982).
- NEW TECHNOLOGICAL SYSTEMS. *Keirstead (1948), in his exposition of a Schumpeterian theory of economic development, introduced the concept of 'constellations' of innovations, which were technically and economically inter-related. Obvious examples are the* *clusters of synthetic materials innovations and petrochemical innovations in the thirties, forties and fifties. They include numerous radical and incremental innovations in both products and processes* (Freeman *et al*., 1982).

- CHANGES OF TECHNO-ECONOMIC PARADIGM (TECHNOLOGICAL REVOLUTIONS). *These are far-reaching and pervasive changes in technology, affecting many (or even all) branches of the economy, as well as giving rise to entirely new sectors. Examples given by Schumpeter were the steam engine and electric power. Characteristic of this type of technical change is that it affects the input cost structure and the conditions of production and distribution for almost every branch of the economy* (Freeman *et al*., 1982).

A change in techno-economic paradigm thus comprises clusters of radical and incremental innovations and embraces several 'new technology systems' (Coccia, 2005). Once a new Techno-economic paradigm has become established throughout the economy it may be described as a 'technological regime'. Nelson and Winter (1982) have also used the concepts of technological regimes and of natural trajectories in technology. Their General natural trajectories' correspond perhaps most closely to 'paradigms'. Bresnahan and Trajtenberg (1995) state that technical progress and growth appear to be driven by a few “General Purpose Technologies” (GPT’s) such as stream engine, electric motor and semiconductors. GPT’s represent the technological regimes or change of techno-economic paradigm and are characterized by pervasiveness, inherent potential for technical improvements, and innovational complementarities, giving rise to increasing returns-to-scale.

Dosi (1982) used the expression 'change of technological paradigm’ and made comparisons with the analogous approach of Kuhn (1962) to ‘scientific revolutions' and paradigm changes in basic science. In these terms 'incremental innovation' along established technological trajectories may be compared with Kuhn's normal science. Whilst strict analogies are out of place the concept of paradigm change has the bringing out the elements of inertia in the system. Whilst there are similarities in all these concepts, the approach of Perez (1985) is the most systematic and has some important distinguishing features in relation to the structural crises of adaptation with which we are concerned. She argues that the development of a new “Techno-economic paradigm” involves a new “best practice” set of rules and customs for designers, engineers, entrepreneurs and managers, which differs respects from the previously prevailing paradigm. Changes of Techno-economic paradigm are based on combinations of radical product, process and organizational innovations. They occur relatively seldom (perhaps twice in a century) but when they occur, they necessitate changes in the institutional and social framework, as well as in most enterprises if their potential is to be fully exploited. They give rise to major changes in the organizational structure of firms, the skill mix and the management style of industry. The Schumpeterian “creative destruction” may thus be only partial in some instances. Note that nothing is said so far about what is being changed, nor in what sense the change is taking place. Nelson and Winter (1982) introduced the concept of technological trajectories to describe both continuous changes and discontinuities in technological innovations: continuous changes are often related to progress along a technological trajectory - the direction of advance within a technological paradigm - while discontinuities are associated with the emergence of a new technological paradigm.

Forces from which technological innovation may originate are two: market pull versus technology push (Darroch and Jardine, 2002). In fact, for this reason innovations are often characterized as incremental versus radical. Dosi (1988) states that an incremental innovation is more likely to be a market pull innovation, while a radical innovation is generally originated by scientists and often incorporates new technologies or new combinations of existing technologies (Van de Ven and Garud, 1993). Thus, radical innovation is often a technology push innovation (Cooper, 1979; Green *et al*., 1995; O’Connor, 1998).

### *1.2 Classifications of innovations within the management of technology (MOT)*

Within the MOT, the classifications of innovation are focused on product, competition, system of production and market. Abernathy and Clark (1985) provide the most important taxonomy. They define transilience as: the significance of innovation for competition depends on its capacity to influence the firm's existing resources, skills and knowledge - What we shall call its 'transilience'. Similarly in Dosi's words (1982; 1988): "Progress upon a technological trajectory is likely to retain some cumulative features". The market transilience scale is in the vertical dimension, and the technology transilience scale in the horizontal. This creates a transilience map, with four quadrants representing a different kind of innovation (Figure 1).

ARCHITECTURAL INNOVATION. New technology that departs from established systems of production, and in turn opens up new linkages to markets and users, is characteristic of the creation of new industries as well as the reformation of old ones. Innovation of this sort defines the basic configuration of product and process, and establishes, the technical and marketing agendas that will guide subsequent development. In effect, it lays down the architecture of the industry, the broad framework within which competition will occur and develop. They have thus labeled innovation of this sort "Architectural"; it is graphed in the upper right hand quadrant of the transilience map.

INNOVATION IN THE MARKET NICHE. Using new concepts in technology to forge new market linkages is the essence of architectural innovation. Opening new market opportunities through the use of existing technology is central to the kind of innovation they have labeled "Niche Creation", but here the effect on production and technical systems is to conserve and strengthen established designs. There are numerous examples of niche creation innovation, ranging from the Timex example referred to earlier, to producers of fashion apparel, and consumer electronics products. The portable radio or cassette player in Sony's Walkman, used established technologies to create a new niche in personal audio products. Innovation of this sort represents what Utterback (1996) has called sales maximization, in which an otherwise stable and well specified technology is refined, improved or changed in a way that supports a new marketing thrust. In some instances, niche creation involves a truly trivial change in technology, in which the impact on productive systems and technical knowledge is incremental. But this type of innovation may also appear in concert with significant new product introductions, vigorous competition on the basis of features, technical refinements, and even technological shifts. The important point is that these changes build on established technical competence, and improve its applicability in emerging market segments.

REGULAR INNOVATION. *The creation of niches and the laying down of a new architecture involve innovation- that is visible and after the fact apparently logical. In contrast, what they call "Regular" innovation is often almost invisible, yet can have a dramatic cumulative effect on product cost and performance. Regular innovation involves change that builds on established technical and production competence and that is applied to existing markets and customers. The effect of these changes is to entrench existing skills and resources. Research on rocket engines, computers and synthetic-fibers has shown that regular innovation can have dramatic effect on production costs, reliability and performance. Regular innovation can have a significant effect on product characteristics and thus can serve to strengthen and entrench not only competence in production, but linkages to customers and markets. It is important to note that these effects tend to take place over a significant period of time. They require an organizational environment and managerial skills that support the dogged pursuit of improvement, no matter how minor. The effects of a given regular innovation on competition are thus of less connect than the cumulative effects of a whole series of changes* (Abernathy and Clark, 1985).

| | | | |
|---|---|---|---|
| | Disrupt Existing/Create New Linkages | | |
| Conserve/Entrecth Existing Competence | Niche Creation | Architectural | Disrupt/Obsolete Existing Competence |
| | Regular | Revolutionary | |
| | Conserve/Entrecth Existing Linkages | | |

*Source*: Abernathy and Clark (1985)

**Fig. 1**: Transilience map

– REVOLUTION INNOVATION. *Innovation that disrupts and renders established technical and production competence obsolete, yet is applied to existing markets and customers, is the fourth category in the transilience map and is labeled "Revolutionary". The reciprocating engine in aircraft, vacuum tubes, and mechanical calculators are recent examples of established technologies that have been over thrown through a revolutionary design. Yet the classic case of revolutionary innovation is the competitive duel between Ford and General Motors in the late 1920s and early 1930s* (Abernathy and Clark, 1985).

A second classical and useful categorization distinguishes between product and process innovation. Utterback (1996) provides the definition of a discontinuous or radical innovation "change that sweeps away much of a firm's existing investment in technical skills and knowledge, designs, production technique, plant and equipment". Rothwell and Gardiner (1998) focus on technological discontinuity: innovations are radically new inventions establishing landmark new products, and as such, create new industries; reinnovations or improvements on existing product design (incremental), existing product (generational), new products (new mark products), improved materials, improving existing products (improvements), improving subsystems of existing products (minor details). Kleinschmidt and Cooper (1991) distinguish between high-moderate-low innovativeness. Anderson and Tushman (1986) actually suggested a typology of technical change that mixes the radical/ incremental categories with the product/process classification. In a more dynamic way Abernathy and Utterback (A-U, 1985) put forward a well-known model also linking these variables and describing patterns of innovations. The A-U model led to the concept that a "dominant technology" emerges, as innovation becomes essentially incremental. The A-U model, although purely descriptive, stands as a landmark in the strategic management literature on technological innovation. Moreover the S-curve has been used to describe the origin and evolution of technologically discontinuous/ radical innovations (Figure 2).

A major problem that arises from A–U's model lies in their binary categorization of the intensity of innovation: are there not indeed innovations that are neither radical nor incremental? Abernathy and Clark (1985) acknowledged this point. They clearly suggest a continuum for technical change defined by polar extremes: their scale ranges from incremental to radical innovation. They do not say however what falls in between and where. There is indeed a continuously varying intensity of change. All radical innovation is not equally radical; all incremental innovation is not just an additional small improvement of what already exists. There are some intermediary changes that both disrupt and continue. The order breaking / order creating distinction suggested by Anderson and Tushman should allow for order breaking-creating categories.

Durand (1992), among these categories, introduces the micro radical innovations. In fact he states that "Moving from 64k DRAM's electronic memories to 128k then to 256k, etc., were by no means radical changes, nor were they simply incremental evolutions".

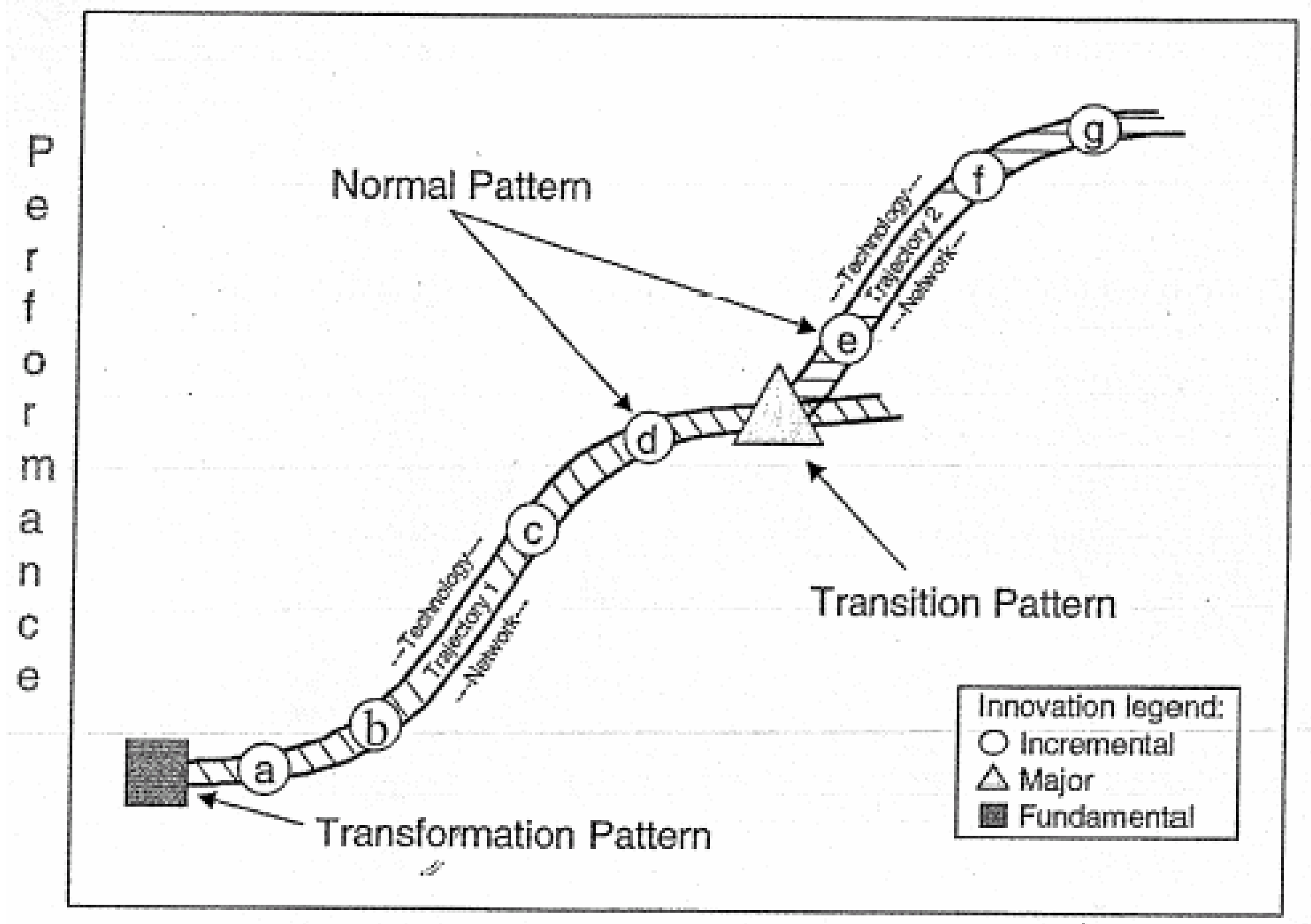

**Fig. 2**: Types of innovation. Source: Rycroft R. W., Kash D. E. (2002)

In synthesis the management of technology uses the following taxonomies in order to classify product innovation (Garcia and Calantone, 2002):

- eight categories – reformulated, new parts, remarchandising, new improvements, new products, new users, new market, new customers (Johnson and Jones, 1957);
- five categories – systematic, major, minor, incremental, unrecorded (Freeman, 1994);
- tetra categorization – incremental, modular, architectural, radical (Henderson and Clark, 1990); niche creation, architectural, regular, revolutionary (Abernathy and Clark, 1985); incremental, evolutionary market, evolutionary technical, radical (Moriarty and Kosnik, 1990); incremental, market breakthrough, technological breakthrough, radical (Chandy and Tellis, 2000); incremental, architectural, fusion, breakthrough (Tidd, 1995; Tidd *et al.* 2001);
- triadic categorization - low, moderate, high innovativeness (Kleinschmidt and Cooper, 1991); incremental, new generation, radically new (Wheelwright and Clark, 1992);
- dichotomous categorization – discontinuous, continuous (Anderson and Tushman, 1990; Robertson, 1967); instrumental, ultimate (Grossman, 1970); variations, reorientation (Norman, 1971); true, adoption (Maidique and Zirger, 1984); original, reformulated (Yoon and Lilien, 1985); innovations, reinnovations (Rothwell and Gardiner, 1988); radical, routine (Meyers and Tucker, 1989); evolutionary, revolutionary (Utterback, 1996); sustaining, disruptive (Christensen, 1997); really new, incremental (Schmidt and Calantone, 1998; Song and Montoya-Weisse, 1998); breakthrough, incremental (Rice *et al.*, 1998); radical, incremental (Balachandra and Friar, 1997; Freeman, 1994).

Garcia and Calantone (2002) use the level macro versus micro, marketing versus technology perspectives and apply Boolean logic to identify three labels for innovations: radical, really new and incremental. The radical innovations are those which cause discontinuity of marketing and technology, both at a macro and a micro level (Van de Ven and Garud, 1993). Incremental innovations occur only at the micro level and cause either discontinuity of marketing, or discontinuity of technology, but not both. Really new innovations include combinations of these two extremes. These three definitions show a reduction in the degree of innovativeness in the following way: radical→ really new → incremental. Moreover, at the macro level, the discontinuities are exogenous to the firm. At both a macro and a micro level the greater the innovativeness -as far as discontinuity of marketing/technology is concerned- the greater the impact on the innovative products. If the discontinuity of the market or of the technology is low, the product will have a low level of innovativeness.

## 2. DISCUSSION AND FUTURE DIRECTIONS

The abundance of types mentioned means that different types of innovation are called by the same name and the same innovation is classified in different manners. Gertrude Stein and William Shakespeare stated “a rose is a rose is a rose. And a rose by any other name would smell just as sweet”. Garcia and Calantone (2002) mention some examples such as the typewriter and the Canon laser photocopier to show that the same innovation can be placed at the beginning or the end of the scale, according to the researcher. The ambiguity of this classification makes it impossible to compare the various studies and, according to the authors, the numerous typologies existing in the economics of innovation, technometrics, economics of technical change, MOT, etc., have hindered the development of knowledge in these fields.

A new taxonomy that tries to overcome the limits of exiting classification it is the scale of technological innovation intensity, elaborated by Coccia (2005). The idea was to develop a way to measures the economic impact of innovation and to provide assessment of the effects on the geo-economic system. In fact, in the economics of innovation, there are no classifications for the effects of innovation on the economic system, although in other fields there are many different scales used to classify (and quantify) an event or the power of a change. Amongst the most common examples is the MCS scale (Mercalli, 1883; Cancani, 1903; Sieberg, 1930) or the Richter scale 1958 used in geophysics to measure the intensity magnitude of earthquakes; the international scale of nuclear events (INES); the scale invented by the English admiral Beaufort to measure the force of the wind, the Douglas scale concerning the state of the sea, the Saffir-Simpson scale for Hurricanes, and so on. For this reason, a scale of technological innovation intensity, is a meta-taxonomy of the economic impact of technological innovation, subsuming less comprehensive taxonomies (Coccia, 2005). This new approach is called ‘seismic’ because the aim is to classify and quantify innovation and technical change through an evaluation scale similar to that used in seismology by Mercalli, who evaluate the intensity of earthquakes through the description of the effects on the geographical environment. In fact, according to this new approach to classify technical change, the socio-economic system is changed by innovations that modify the economic space with a series of effects both on the subjects and the objects (the technological intensity measures the strength of the technical change produced by the innovation within the economy). Moreover, the metrics of this approach quantifies the economic and social impact of technical change, over time and space, through a indicator called Magnitude of Technical Change (Figure 3).

The scale of innovation intensity is described in table 1. It shows that some of the taxonomies for innovations presented in economic literature are synthesized in new levels, called innovation degrees. This scale synthesizes the abundance of innovation taxonomies presented in economic literature where different types of innovation are called by the same name and same innovations are classified in different levels. The innovation degree allows a comparison among various innovations, which is impossible with the previous classifications.

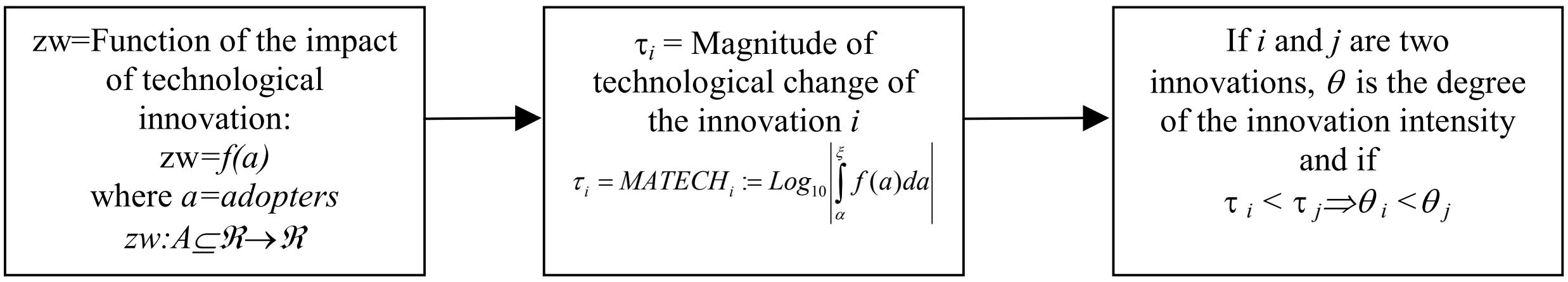

**Fig. 3**: Steps to measure the intensity of technological change according to the seismic approach

**Tab. 1**: Scale of innovation intensity

| CLASSIFICATION OF TECHNICAL CHANGE | | | |
|---|---|---|---|
| *Economic impact* | *Innovation Degree* | *Innovation Intensity* | *Some definitions of innovations used in economic literature* |
| I SET<br>Low Impact | I=1st | Lightest | Elementary or micro-incremental (Coccia, 2005)<br>Unrecorded (Freeman, 1994) |
| | II=2nd | Mild | Continuos (Freeman et *al.*, 1982)<br>Improvements (Mensch, 1979)<br>Incremental (Freeman *et al*., 1982; Priest and Hill, 1980)<br>Market Pull (Dosi, 1988)<br>Minor (Archibugi and Santarelli, 1989)<br>Normal science (Khun, 1962)<br>Regular (Abernathy and Clark, 1985) |
| | III=3rd | Moderate | Major (Archibugi and Santarelli, 1989; Rycroft and Kash, 2002)<br>Market breakthrough (Chandy and Tellis, 2000)<br>Modular (Henderson and Clark, 1990)<br>Non drastic (Arrow, 1962); Gilbert and Newbery, 1982)<br>Really new (Garcia and Calantone, 2002) |
| II SET<br>Medium Impact | IV=4th | Intermediate | Evolutionary Technical (Moriarty and Kosnik, 1990)<br>Micro-radical (Durand, 1992)<br>Niche creation (Abernathy and Clark, 1985)<br>Non drastic (Arrow, 1962; Gilbert and Newbery, 1982)<br>Technological breakthrough (Chandy and Tellis, 2000) |
| | V=5th | Strong | Architectural (Abernathy and Clark, 1985)<br>Basic innovation (Mensch, 1979)<br>Breakthrough (Tidd, 1995)<br>Discontinuous (Archibugi and Santarelli, 1989)<br>Discrete (Priest and Hill, 1980)<br>Drastic (Arrow, 1962; Gilbert and Newbery, 1982)<br>Fundamental (Mensch, 1979)<br>Radical (Freeman *et al*., 1982)<br>Technology push (Dosi, 1988) |
| III SET<br>High Impact | VI=6th | Very strong | Clusters of innovations (Freeman *et al*., 1982)<br>Constellations of innovations (Keirstead, 1948)<br>Innovation systems (Sahal, 1981)<br>New technological Systems (Freeman *et al*., 1982) |
| | VII=7th | Revolutionary | Change of Techno-economic paradigms (Freeman *et al*., 1982)<br>Change of technological paradigms (Dosi, 1982)<br>Cluster of New technological systems (Coccia, 2005)<br>Revolutionary (Abernathy and Clark, 1985)<br>Technological regimes (Nelson and Winter, 1982)<br>Technological revolutions (Freeman *et al*., 1982; Freeman, 1984) |

Coccia (2005) argues that although the causes that originate innovations are different, the effects on the geo-economics systems are always similar. For instance the effects produced by the VII degree innovations are the following:

- The means of human communication are radically changed and a new means of communication, that heavily affects all the economic subjects and objects, has origin, forcing all those who use it to change their habits. A new Techno-economic paradigm is born. The innovation of VII degree produces the following effects on Consumer, Firm and Market.
- *Consumer*. Changes in lifestyle, in habits tendency to save and invest, etc. The standard of living and well being are considerably improved.
- *Firm*. Involves all the firms within the economic system. The organizational structures and production processes are changed. New firms offering new services and/or products are founded. Production, organizational and management methods change.
- *Market*. This innovation revolutionizes all the existing markets, creating new ones. The markets become increasingly turbulent.

## 3. CONCLUDING REMARKS

Nowadays, technical change and innovation play more and more a fundamental role in various fields. Technical change can display in several types of innovation which have different intensity. The classifications and measurements of technical change are important indicators for the economic growth, the consumers' behavior, the analysis of international trade and the evaluation of monetary and fiscal policies. The measurement of technology is the key to the forecasting and management of product and process innovations.

The economic literature on technical change uses different names to indicate the same type of technical change (Archibugi, Freeman, Pavitt, Durand, Abernathy, Clark and others). This diversity is not considered a heterogeny (different elements that make up the innovation) but rather a heterophylly (differing forms of innovation with common origin and genes). The latter generates different definitions of innovations, which are substantially similar, but differ in the form. From this survey I conclude that there is a relatively large number of individual and unconnected innovation classifications that tend to be disconnected from each other or ignore the different input and process issues responsible innovation diversity. For this reason the new directions to classify the technical change are based on common denominators, such as the innovation intensity degree that is based on the effects of innovation on the geo-economic system (subjects, such as consumers and firms; objects such as: means of communication, infrastructures, etc). In general the measurement and classifications of technology is carried out after that the innovative event has occurred. This logic has limits for technological forecasting and foresight but to measure innovation – a posteriori - can be useful, in similar way is useful to measures and classifies the intensity of earthquakes (one of the most unpredictable events) after they have occurred. In fact, a country can learn from past innovations and equip itself with modern infrastructures, means of communication, trained human resources, which are its strength in absorbing and accepting the impact of future technological innovations. This explains why a country with a large number of computers, modern telecommunication networks, universities, modern means of communication, science parks and trained human resources is advantaged in the absorption of new technological innovations and has a competitive advantage in comparison to countries with less technological infrastructures and resources. Moreover all taxonomies present a level called revolution, which indicate a high impact of innovation. For instance, within the scale of technological intensity the intensity (Coccia, 2005) of the $7^{th}$ degree innovation is strictly limited to those innovations that change human communication and have mass diffusion. Seventh degree is similar to the technological revolution of Freeman and Soete (1987) and

revolution – innovation of Abernathy and Clark (1985).

In all, the difficulty presents to classify and measure the innovations is due to several variables of the technical change function. This is a serious problem for all economists and scholars since the measurement and classifications of technical change and innovation cannot be traced to a single discipline, but these difficulties represent a challenge still to be tacked.

WORKING PAPER SERIES (2006-1993)

**2006**

1/06 *Analisi della crescita economica regionale e convergenza: un nuovo approccio teorico ed evidenza empirica sull'Italia*, by Mario Coccia

2/06 *Classifications of innovations: Survey and future directions*, by Mario Coccia

3/06 *Analisi economica dell'impatto tecnologico*, by Mario Coccia

4/06 *La burocrazia nella ricerca pubblica.* PARTE I *Una rassegna dei principali studi*, by Mario Coccia and Alessandro Gobbino

5/06 *La burocrazia nella ricerca pubblica.* PARTE II *Analisi della burocrazia negli Enti Pubblici di Ricerca*, by Mario Coccia and Alessandro Gobbino

6/06 *La burocrazia nella ricerca pubblica.* PARTE III *Organizzazione e Project Management negli Enti Pubblici di Ricerca: l'analisi del CNR*, by Mario Coccia, Secondo Rolfo and Alessandro Gobbino

7/06 *Economic and social studies of scientific research: nature and origins*, by Mario Coccia

8/06 *Shareholder Protection and the Cost of Capital: Empirical Evidence from German and Italian Firms*, by Julie Ann Elston and Laura Rondi

9/06 *Réflexions en thème de district, clusters, réseaux: le problème de la gouvernance*, by Secondo Rolfo

10/06 *Models for Default Risk Analysis: Focus on Artificial Neural Networks, Model Comparisons, Hybrid Frameworks*, by Greta Falavigna

11/06 *Le politiche del governo federale statunitense nell'edilizia residenziale. Suggerimenti per il modello italiano*, by Davide Michelis

12/06 *Il finanziamento delle imprese Spin-off: un confronto fra Italia e Regno Unito*, by Elisa Salvador

13/06 SERIE SPECIALE IN COLLABORAZIONE CON HERMES: *Regulatory and Environmental Effects on Public Transit Efficiency: a Mixed DEA-SFA Approach*, by Beniamina Buzzo Margari, Fabrizio Erbetta, Carmelo Petraglia, Massimiliano Piacenza

14/06 *La mission manageriale: risorsa delle aziende*, by Gian Franco Corio

15/06 *Peer review for the evaluation of the academic research: the Italian experience*, by Emanuela Reale, Anna Barbara, Antonio Costantini

**2005**

1/05 *Gli approcci biologici nell'economia dell'innovazione*, by Mario Coccia

2/05 *Sistema informativo sulle strutture operanti nel settore delle biotecnologie in Italia*, by Edoardo Lorenzetti, Francesco Lutman, Mauro Mallone

3/05 *Analysis of the Resource Concentration on Size and Research Performance. The Case of Italian National Research Council over the Period 2000-2004*, by Mario Coccia and Secondo Rolfo

4/05 *Le risorse pubbliche per la ricerca scientifica e lo sviluppo sperimentale nel 2002*, by Anna Maria Scarda

5/05 *La customer satisfaction dell'URP del Cnr. I casi Lazio, Piemonte e Sicilia*, by Gian Franco Corio

6/05 *La comunicazione integrata tra uffici per le relazioni con il pubblico della Pubblica Amministrazione*, by Gian Franco Corio

7/05 *Un'analisi teorica sul marketing territoriale. Presentazione di un caso studio. Il "consorzio per la tutela dell'Asti"*, by Maria Marenna

8/05 *Una proposta di marketing territoriale: una possibile griglia di analisi delle risorse*, by Gian Franco Corio

9/05 *Analisi e valutazione delle performance economico-tecnologiche di diversi paesi e situazione italiana*, by Mario Coccia and Mario Taretto

10/05 *The patenting regime in the Italian public research system: what motivates public inventors to patent,* by Bianca Potì and Emanuela Reale

11/05 *Changing patterns in the steering of the University in Italy: funding rules and doctoral programmes,* by Bianca Potì and Emanuela Reale

12/05 *Una "discussione in rete" con Stanley Wilder*, by Carla Basili

13/05 *New Tools for the Governance of the Academic Research in Italy: the Role of Research Evaluation*, by Bianca Potì and Emanuela Reale

14/05 *Product Differentiation, Industry Concentration and Market Share Turbulence*, by Catherine Matraves, Laura Rondi

15/05 *Riforme del Servizio Sanitario Nazionale e dinamica dell'efficienza ospedaliera in Piemonte*, by Chiara Canta, Massimiliano Piacenza, Gilberto Turati

16/05 SERIE SPECIALE IN COLLABORAZIONE CON HERMES: *Struttura di costo e rendimenti di scala nelle imprese di trasporto pubblico locale di medie-grandi dimensioni*, by Carlo Cambini, Ivana Paniccia, Massimiliano Piacenza, Davide Vannoni 17/05 *Ricerc@.it - Sistema informativo su istituzioni, enti e strutture di ricerca in Italia*, by Edoardo Lorenzetti, Alberto Paparello

**2004**

1/04 *Le origini dell'economia dell'innovazione: il contributo di Rae*, by Mario Coccia

2/04 *Liberalizzazione e integrazione verticale delle utility elettriche: evidenza empirica da un campione italiano di imprese pubbliche locali*, by Massimiliano Piacenza and Elena Beccio

3/04 *Uno studio sull'innovazione nell'industria chimica*, by Anna Ceci, Mario De Marchi, Maurizio Rocchi

4/04 *Labour market rigidity and firms' R&D strategies*, by Mario De Marchi and Maurizio Rocchi

5/04 *Analisi della tecnologia e approcci alla sua misurazione*, by Mario Coccia

6/04 *Analisi delle strutture pubbliche di ricerca scientifica: tassonomia e comportamento strategico*, by Mario Coccia

7/04 *Ricerca teorica vs. ricerca applicata. Un'analisi relativa al Cnr*, by Mario Coccia and Secondo Rolfo

8/04 *Considerazioni teoriche sulla diffusione delle innovazioni nei distretti industriali: il caso delle ICT*, by Arianna Miglietta

9/04 *Le politiche industriali regionali nel Regno Unito*, by Elisa Salvador

10/04 *Going public to grow? Evidence from a panel of Italian firms*, by Robert E. Carpenter and L. Rondi

11/04 *What Drives Market Prices in the Wine Industry? Estimation of a Hedonic Model for Italian Premium Wine*, by Luigi Benfratello, Massimiliano Piacenza and Stefano Sacchetto

12/04 *Brief notes on the policies for science-based firms*, by Mario De Marchi, Maurizio Rocchi

13/04 *Countrymetrics e valutazione della performance economica dei paesi: un approccio sistemico*, by Mario Coccia

14/04 *Analisi del rischio paese e sistemazione tassonomica*, by Mario Coccia

15/04 *Organizing the Offices for Technology Transfer*, by Chiara Franzoni

16/04 *Le relazioni tra ricerca pubblica e industria in Italia*, by Secondo Rolfo

17/04 *Modelli di analisi e previsione del rischio di insolvenza: una prospettiva delle metodologie applicate*, by Nadia D'Annunzio e Greta Falavigna

18/04 *SERIE SPECIALE: Lo stato di salute del sistema industriale piemontese: analisi economico-finanziaria delle imprese piemontesi*, Terzo Rapporto 1999-2002, by Giuseppe Calabrese, Fabrizio Erbetta, Federico Bruno Rolle

19/04 *SERIE SPECIALE: Osservatorio sulla dinamica economico-finanziaria delle imprese della filiera del tessile e dell'abbigliamento in Piemonte*, Primo rapporto 1999-2002, by Giuseppe Calabrese, Fabrizio Erbetta, Federico Bruno Rolle

20/04 *SERIE SPECIALE: Osservatorio sulla dinamica economico-finanziaria delle imprese della filiera dell'auto in Piemonte,* Secondo Rapporto 1999-2002, by Giuseppe Calabrese, Fabrizio Erbetta, Federico Bruno Rolle

**2003**

1/03 *Models for Measuring the Research Performance and Management of the Public Labs,* by Mario Coccia, March

2/03 *An Approach to the Measurement of Technological Change Based on the Intensity of Innovation,* by Mario Coccia, April

3/03 *Verso una patente europea dell'informazione: il progetto EnIL*, by Carla Basili, June

4/03 *Scala della magnitudo innovativa per misurare l'attrazione spaziale del trasferimento tecnologico,* by Mario Coccia, June

5/03 *Mappe cognitive per analizzare i processi di creazione e diffusione della conoscenza negli Istituti di ricerca*, by Emanuele Cadario, July

6/03 *Il servizio postale: caratteristiche di mercato e possibilità di liberalizzazione*, by Daniela Boetti, July

7/03 *Donne-scienza-tecnologia: analisi di un caso di studio*, by Anita Calcatelli, Mario Coccia, Katia Ferraris and Ivana Tagliafico, July

8/03 SERIE SPECIALE. OSSERVATORIO SULLE PICCOLE IMPRESE INNOVATIVE TRIESTE. *Imprese innovative in Friuli Venezia Giulia: un esperimento di analisi congiunta*, by Lucia Rotaris, July

9/03 *Regional Industrial Policies in Germany*, by Helmut Karl*,* Antje Möller and Rüdiger Wink, July

10/03 SERIE SPECIALE. OSSERVATORIO SULLE PICCOLE IMPRESE INNOVATIVE TRIESTE. *L'innovazione nelle new technology-based firms in Friuli-Venezia Giulia*, by Paola Guerra, October

11/03 SERIE SPECIALE. *Lo stato di salute del sistema industriale piemontese: analisi economico-finanziaria delle imprese piemontesi,* Secondo Rapporto 1998-2001, December

12/03 SERIE SPECIALE. *Osservatorio sulla dinamica economico-finanziaria delle imprese della meccanica specializzata in Piemonte*, Primo Rapporto 1998-2001, December

13/03 SERIE SPECIALE. *Osservatorio sulla dinamica economico-finanziaria delle imprese delle bevande in Piemonte*, Primo Rapporto 1998-2001, December

**2002**

1/02 *La valutazione dell'intensità del cambiamento tecnologico: la scala Mercalli per le innovazioni*, by Mario Coccia, January

2/02 SERIE SPECIALE IN COLLABORAZIONE CON HERMES. *Regulatory constraints and cost efficiency of the Italian public transit systems: an exploratory stochastic frontier model*, by Massimiliano Piacenza, March

3/02 *Aspetti gestionali e analisi dell'efficienza nel settore della distribuzione del gas*, by Giovanni Fraquelli and Fabrizio Erbetta, March

4/02 *Dinamica e comportamento spaziale del trasferimento tecnologico*, by Mario Coccia, April

5/02 *Dimensione organizzativa e performance della ricerca: l'analisi del Consiglio Nazionale delle Ricerche*, by Mario Coccia and Secondo Rolfo, April

6/02 *Analisi di un sistema innovativo regionale e implicazioni di policy nel processo di trasferimento tecnologico,* by Monica Cariola and Mario Coccia, April

7/02 *Analisi psico-economica di un'organizzazione scientifica e implicazioni di management: l'Istituto Elettrotecnico Nazionale "G. Ferraris"*, by Mario Coccia and Alessandra Monticone, April

8/02 *Firm Diversification in the European Union. New Insights on Return to Core Business and Relatedness*, by Laura Rondi and Davide Vannoni, May

9/02 *Le nuove tecnologie di informazione e comunicazione nelle PMI: un'analisi sulla diffusione dei siti internet nel distretto di Biella,* by Simona Salinari, June

10/02 *La valutazione della soddisfazione di operatori di aziende sanitarie,* by Gian Franco Corio, November

11/02 *Analisi del processo innovativo nelle PMI italiane*, by Giuseppe Calabrese, Mario Coccia and Secondo Rolfo, November

12/02 *Metrics della Performance dei laboratori pubblici di ricerca e comportamento strategico*, by Mario Coccia, September

13/02 *Technometrics basata sull'impatto economico del cambiamento tecnologico*, by Mario Coccia, November

**2001**

1/01 *Competitività e divari di efficienza nell'industria italiana*, by Giovanni Fraquelli, Piercarlo Frigero and Fulvio Sugliano, January

2/01 *Waste water purification in Italy: costs and structure of the technology*, by Giovanni Fraquelli and Roberto Giandrone, January

3/01 SERIE SPECIALE IN COLLABORAZIONE CON HERMES. *Il trasporto pubblico locale in Italia: variabili esplicative dei divari di costo tra le imprese*, by Giovanni Fraquelli, Massimiliano Piacenza and Graziano Abrate, February

4/01 *Relatedness, Coherence, and Coherence Dynamics: Empirical Evidence from Italian Manufacturing*, by Stefano Valvano and Davide Vannoni, February

5/01 *Il nuovo panel Ceris su dati di impresa 1977-1997,* by Luigi Benfratello, Diego Margon, Laura Rondi, Alessandro Sembenelli, Davide Vannoni, Silvana Zelli, Maria Zittino, October

6/01 *SMEs and innovation: the role of the industrial policy in Italy*, by Giuseppe Calabrese and Secondo Rolfo, May

7/01 *Le martingale: aspetti teorici ed applicativi*, by Fabrizio Erbetta and Luca Agnello, September

8/01 *Prime valutazioni qualitative sulle politiche per la R&S in alcune regioni italiane*, by Elisa Salvador, October

9/01 *Accords technology transfer-based: théorie et méthodologie d'analyse du processus*, *by Mario Coccia,* October

10/01 *Trasferimento tecnologico: indicatori spaziali,* by Mario Coccia, November

11/01 *Does the run-up of privatisation work as an effective incentive mechanism? Preliminary findings from a sample of Italian firms*, by Fabrizio Erbetta, October

12/01 SERIE SPECIALE IN COLLABORAZIONE CON HERMES. *Costs and Technology of Public Transit Systems in Italy: Some Insights to Face Inefficiency*, by Giovanni Fraquelli, Massimiliano Piacenza and Graziano Abrate, October

13/01 *Le NTBFs a Sophia Antipolis, analisi di un campione di imprese*, by Alessandra Ressico, December

**2000**

1/00 *Trasferimento tecnologico: analisi spaziale,* by Mario Coccia, March

2/00 *Poli produttivi e sviluppo locale: una indagine sulle tecnologie alimentari nel mezzogiorno,* by Francesco G. Leone, March

3/00 *La mission del top management di aziende sanitarie,* by Gian Franco Corio, March

4/00 *La percezione dei fattori di qualità in Istituti di ricerca: una prima elaborazione del caso Piemonte,* by Gian Franco Corio, March

5/00 *Una metodologia per misurare la performance endogena nelle strutture di R&S,* by Mario Coccia, April

6/00 *Soddisfazione, coinvolgimento lavorativo e performance della ricerca,* by Mario Coccia, May

7/00 *Foreign Direct Investment and Trade in the EU: Are They Complementary or Substitute in Business Cycles Fluctuations?*, by Giovanna Segre, April

8/00 *L'attesa della privatizzazione: una minaccia credibile per il manager?*, by Giovanni Fraquelli, May

9/00 *Gli effetti occupazionali dell'innovazione. Verifica su un campione di imprese manifatturiere italiane*, by Marina Di Giacomo, May 10/00 *Investment, Cash Flow and Managerial Discretion in State-owned Firms. Evidence Across Soft and Hard Budget Constraints*, by Elisabetta Bertero and Laura Rondi, June
11/00 *Effetti delle fusioni e acquisizioni: una rassegna critica dell'evidenza empirica*, by Luigi Benfratello, June
12/00 *Identità e immagine organizzativa negli Istituti CNR del Piemonte*, by Paolo Enria, August
13/00 *Multinational Firms in Italy: Trends in the Manufacturing Sector*, by Giovanna Segre, September
14/00 *Italian Corporate Governance, Investment, and Finance*, by Robert E. Carpenter and Laura Rondi, October
15/00 *Multinational Strategies and Outward-Processing Trade between Italy and the CEECs: The Case of Textile-Clothing*, by Giovanni Balcet and Giampaolo Vitali, December
16/00 *The Public Transit Systems in Italy: A Critical Analysis of the Regulatory Framework*, by Massimiliano Piacenza, December

**1999**
1/99 *La valutazione delle politiche locali per l'innovazione: il caso dei Centri Servizi in Italia,* by Monica Cariola and Secondo Rolfo, January
2/99 *Trasferimento tecnologico ed autofinanziamento: il caso degli Istituti Cnr in Piemonte,* by Mario Coccia, March
3/99 *Empirical studies of vertical integration: the transaction cost orthodoxy,* by Davide Vannoni, March
4/99 *Developing innovation in small-medium suppliers: evidence from the Italian car industry,* by Giuseppe Calabrese, April
5/99 *Privatization in Italy: an analysis of factors productivity and technical efficiency,* by Giovanni Fraquelli and Fabrizio Erbetta, March
6/99 *New Technology Based-Firms in Italia: analisi di un campione di imprese triestine,* by Anna Maria Gimigliano, April
7/99 *Trasferimento tacito della conoscenza: gli Istituti CNR dell'Area di Ricerca di Torino,* by Mario Coccia, May
8/99 *Struttura ed evoluzione di un distretto industriale piemontese: la produzione di casalinghi nel Cusio,* by Alessandra Ressico, June
9/99 *Analisi sistemica della performance nelle strutture di ricerca,* by Mario Coccia, September
10/99 *The entry mode choice of EU leading companies (1987-1997),* by Giampaolo Vitali, November
11/99 *Esperimenti di trasferimento tecnologico alle piccole e medie imprese nella Regione Piemonte,* by Mario Coccia, November
12/99 *A mathematical model for performance evaluation in the R&D laboratories: theory and application in Italy,* by Mario Coccia, November
13/99 *Trasferimento tecnologico: analisi dei fruitori,* by Mario Coccia, December
14/99 *Beyond profitability: effects of acquisitions on technical efficiency and productivity in the Italian pasta industry,* by Luigi Benfratello, December
15/99 *Determinanti ed effetti delle fusioni e acquisizioni: un'analisi sulla base delle notifiche alle autorità antitrust,* by Luigi Benfratello, December

**1998**
1/98 *Alcune riflessioni preliminari sul mercato degli strumenti multimediali,* by Paolo Vaglio, January
2/98 *Before and after privatization: a comparison between competitive firms,* by Giovanni Fraquelli and Paola Fabbri, January
3/98 **Not available**
4/98 *Le importazioni come incentivo alla concorrenza: l'evidenza empirica internazionale e il caso del mercato unico europeo,* by Anna Bottasso, May
5/98 *SEM and the changing structure of EU Manufacturing, 1987-1993,* by Stephen Davies, Laura Rondi and Alessandro Sembenelli, November
6/98 *The diversified firm: non formal theories versus formal models,* by Davide Vannoni, December
7/98 *Managerial discretion and investment decisions of state-owned firms: evidence from a panel of Italian companies,* by Elisabetta Bertero and Laura Rondi, December
8/98 *La valutazione della R&S in Italia: rassegna delle esperienze del C.N.R. e proposta di un approccio alternativo,* by Domiziano Boschi, December
9/98 *Multidimensional Performance in Telecommunications, Regulation and Competition: Analysing the European Major Players,* by Giovanni Fraquelli and Davide Vannoni, December

**1997**
1/97 *Multinationality, diversification and firm size. An empirical analysis of Europe's leading firms,* by Stephen Davies, Laura Rondi and Alessandro Sembenelli, January
2/97 *Qualità totale e organizzazione del lavoro nelle aziende sanitarie,* by Gian Franco Corio, January
3/97 *Reorganising the product and process development in Fiat Auto,* by Giuseppe Calabrese, February
4/97 *Buyer-supplier best practices in product development: evidence from car industry,* by Giuseppe Calabrese, April

5/97 *L'innovazione nei distretti industriali. Una rassegna ragionata della letteratura,* by Elena Ragazzi, April

6/97 *The impact of financing constraints on markups: theory and evidence from Italian firm level data,* by Anna Bottasso, Marzio Galeotti and Alessandro Sembenelli, April

7/97 *Capacità competitiva e evoluzione strutturale dei settori di specializzazione: il caso delle macchine per confezionamento e imballaggio,* by Secondo Rolfo, Paolo Vaglio, April

8/97 *Tecnologia e produttività delle aziende elettriche municipalizzate,* by Giovanni Fraquelli and Piercarlo Frigero, April

9/97 *La normativa nazionale e regionale per l'innovazione e la qualità nelle piccole e medie imprese: leggi, risorse, risultati e nuovi strumenti,* by Giuseppe Calabrese, June

10/97 *European integration and leading firms' entry and exit strategies,* by Steve Davies, Laura Rondi and Alessandro Sembenelli, April

11/97 *Does debt discipline state-owned firms? Evidence from a panel of Italian firms,* by Elisabetta Bertero and Laura Rondi, July

12/97 *Distretti industriali e innovazione: i limiti dei sistemi tecnologici locali,* by Secondo Rolfo and Giampaolo Vitali, July

13/97 *Costs, technology and ownership form of natural gas distribution in Italy,* by Giovanni Fraquelli and Roberto Giandrone, July

14/97 *Costs and structure of technology in the Italian water industry,* by Paola Fabbri and Giovanni Fraquelli, July

15/97 *Aspetti e misure della customer satisfaction/dissatisfaction,* by Maria Teresa Morana, July

16/97 *La qualità nei servizi pubblici: limiti della normativa UNI EN 29000 nel settore sanitario,* by Efisio Ibba, July

17/97 *Investimenti, fattori finanziari e ciclo economico,* by Laura Rondi and Alessandro Sembenelli, rivisto sett. 1998

18/97 *Strategie di crescita esterna delle imprese leader in Europa: risultati preliminari dell'utilizzo del data-base Ceris "100 top EU firms' acquisition/divestment database 1987-1993",* by Giampaolo Vitali and Marco Orecchia, December

19/97 *Struttura e attività dei Centri Servizi all'innovazione: vantaggi e limiti dell'esperienza italiana,* by Monica Cariola, December

20/97 *Il comportamento ciclico dei margini di profitto in presenza di mercati del capitale meno che perfetti: un'analisi empirica su dati di impresa in Italia,* by Anna Bottasso, December

**1996**

1/96 *Aspetti e misure della produttività. Un'analisi statistica su tre aziende elettriche europee,* by Donatella Cangialosi, February

2/96 *L'analisi e la valutazione della soddisfazione degli utenti interni: un'applicazione nell'ambito dei servizi sanitari,* by Maria Teresa Morana, February

3/96 *La funzione di costo nel servizio idrico. Un contributo al dibattito sul metodo normalizzato per la determinazione della tariffa del servizio idrico integrato,* by Giovanni Fraquelli and Paola Fabbri, February

4/96 *Coerenza d'impresa e diversificazione settoriale: un'applicazione alle società leaders nell'industria manifatturiera europea,* by Marco Orecchia, February

5/96 *Privatizzazioni: meccanismi di collocamento e assetti proprietari. Il caso STET,* by Paola Fabbri, February

6/96 *I nuovi scenari competitivi nell'industria delle telecomunicazioni: le principali esperienze internazionali,* by Paola Fabbri, February

7/96 *Accordi, joint-venture e investimenti diretti dell'industria italiana nella CSI: Un'analisi qualitativa,* by Chiara Monti and Giampaolo Vitali, February

8/96 *Verso la riconversione di settori utilizzatori di amianto. Risultati di un'indagine sul campo,* by Marisa Gerbi Sethi, Salvatore Marino and Maria Zittino, February

9/96 *Innovazione tecnologica e competitività internazionale: quale futuro per i distretti e le economie locali,* by Secondo Rolfo, March

10/96 *Dati disaggregati e analisi della struttura industriale: la matrice europea delle quote di mercato,* by Laura Rondi, March

11/96 *Le decisioni di entrata e di uscita: evidenze empiriche sui maggiori gruppi italiani,* by Alessandro Sembenelli and Davide Vannoni, April

12/96 *Le direttrici della diversificazione nella grande industria italiana,* by Davide Vannoni, April

13/96 *R&S cooperativa e non-cooperativa in un duopolio misto con spillovers,* by Marco Orecchia, May

14/96 *Unità di studio sulle strategie di crescita esterna delle imprese italiane*, by Giampaolo Vitali and Maria Zittino, July. **Not available**

15/96 *Uno strumento di politica per l'innovazione: la prospezione tecnologica,* by Secondo Rolfo, September

16/96 *L'introduzione della Qualità Totale in aziende ospedaliere: aspettative ed opinioni del middle management,* by Gian Franco Corio, September 17/96 *Shareholders' voting power and block transaction premia: an empirical analysis of Italian listed companies,* by Giovanna Nicodano and Alessandro Sembenelli, November
18/96 *La valutazione dell'impatto delle politiche tecnologiche: un'analisi classificatoria e una rassegna di alcune esperienze europee,* by Domiziano Boschi, November
19/96 *L'industria orafa italiana: lo sviluppo del settore punta sulle esportazioni,* by Anna Maria Gaibisso and Elena Ragazzi, November
20/96 *La centralità dell'innovazione nell'intervento pubblico nazionale e regionale in Germania,* by Secondo Rolfo, December
21/96 *Ricerca, innovazione e mercato: la nuova politica del Regno Unito,* by Secondo Rolfo, December
22/96 *Politiche per l'innovazione in Francia,* by Elena Ragazzi, December
23/96 *La relazione tra struttura finanziaria e decisioni reali delle imprese: una rassegna critica dell'evidenza empirica,* by Anna Bottasso, December

**1995**

1/95 *Form of ownership and financial constraints: panel data evidence on leverage and investment choices by Italian firms,* by Fabio Schiantarelli and Alessandro Sembenelli, March
2/95 *Regulation of the electric supply industry in Italy,* by Giovanni Fraquelli and Elena Ragazzi, March
3/95 *Restructuring product development and production networks: Fiat Auto,* by Giuseppe Calabrese, September
4/95 *Explaining corporate structure: the MD matrix, product differentiation and size of market,* by Stephen Davies, Laura Rondi and Alessandro Sembenelli, November
5/95 *Regulation and total productivity performance in electricity: a comparison between Italy, Germany and France,* by Giovanni Fraquelli and Davide Vannoni, December
6/95 *Strategie di crescita esterna nel sistema bancario italiano: un'analisi empirica 1987-1994,* by Stefano Olivero and Giampaolo Vitali, December
7/95 *Panel Ceris su dati di impresa: aspetti metodologici e istruzioni per l'uso,* by Diego Margon, Alessandro Sembenelli and Davide Vannoni, December

**1994**

1/94 *Una politica industriale per gli investimenti esteri in Italia: alcune riflessioni,* by Giampaolo Vitali, May
2/94 *Scelte cooperative in attività di ricerca e sviluppo,* by Marco Orecchia, May
3/94 *Perché le matrici intersettoriali per misurare l'integrazione verticale?,* by Davide Vannoni, July
4/94 *Fiat Auto: A simultaneous engineering experience,* by Giuseppe Calabrese, August

**1993**

1/93 *Spanish machine tool industry,* by Giuseppe Calabrese, November
2/93 *The machine tool industry in Japan,* by Giampaolo Vitali, November
3/93 *The UK machine tool industry,* by Alessandro Sembenelli and Paul Simpson, November
4/93 *The Italian machine tool industry,* by Secondo Rolfo, November
5/93 *Firms' financial and real responses to business cycle shocks and monetary tightening: evidence for large and small Italian companies,* by Laura Rondi, Brian Sack, Fabio Schiantarelli and Alessandro Sembenelli, December